\documentclass{article}

\usepackage{arxiv}

\usepackage[utf8]{inputenc} 
\usepackage[T1]{fontenc}    
\usepackage{hyperref}       
\usepackage{url}            
\usepackage{booktabs}       
\usepackage{amsfonts}       
\usepackage{nicefrac}       
\usepackage{microtype}      
\usepackage{lipsum}		
\usepackage{doi}
\usepackage{array}
\usepackage{enumitem}
\usepackage{graphicx}
\usepackage[square,sort,comma,numbers]{natbib}

\title{The Office of Astronomy for Development Impact Cycle}

\author{ \href{https://orcid.org/0000-0002-9745-0504}{\includegraphics[scale=0.06]{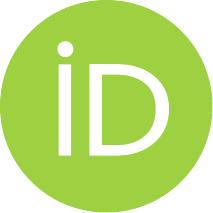}\hspace{1mm}Joyful E. Mdhluli}{ on behalf of the IAU Office of Astronomy for Development} \thanks{Visit our website, www.astro4dev.org or email: info@astro4dev.org} \\
	International Astronomical Union's Office of Astronomy for Development\\
	South African Astronomical Observatory\\
        Cape Town, South Africa\\
	\texttt{joy@astro4dev.org} \\
	\And
}

\renewcommand{\shorttitle}{\it{OAD Impact Cycle}}

\hypersetup{
pdftitle={The Office of Astronomy for Development Impact Cycle},
pdfsubject={astro-ph.IM},
pdfauthor={Joyful E.~Mdhluli},
pdfkeywords={Astronomy, Sustainable development, Impact},
}

\begin{document}
\maketitle

\begin{abstract}
The Office of Astronomy for Development (OAD) believes that in order for astronomy-for-development activities to be effective, a scientific approach is required. Evaluation is an essential component in identifying which projects work best, for whom and under what conditions. Evidence-informed project design and selection ensures that projects build on past lessons, thereby reducing the risk of negative unintended consequences and increasing the probabilities of positive cost-effective impact. The OAD has developed an Impact Cycle that aims to enhance project design, selection and delivery systems to support such continual improvement and potential expansion. By determining what works - and, importantly, what doesn’t work - the OAD can build a library of evidence on best practice and ensure a positive feedback loop for future projects.
\end{abstract}


\section{Introduction}
The global Office of Astronomy for Development (OAD) is a joint partnership between the International Astronomical Union (IAU) and the South African National Research Foundation (NRF) with the support of the Department of Science, Technology and Innovation (DSTI). The mission of the OAD is to further the use of astronomy, in all its aspects, as a tool for development.

One of the primary ways the OAD implements its mission is through the Call for Proposals, which have been conducted annually since 2012. Every year, the OAD invites proposals for projects that use astronomy as a tool to address one or more challenges related to sustainable development. The call for proposals is open to anyone from anywhere in the world\footnote{Due to international sanctions, we are unable to send money to certain countries.}.  The OAD Call for Proposals is conducted in two stages, where only a limited number of proposals from Stage 1 will be invited to submit a Stage 2 proposal.

\begin{figure}[h!]
    \centering
    \includegraphics[width=\textwidth]{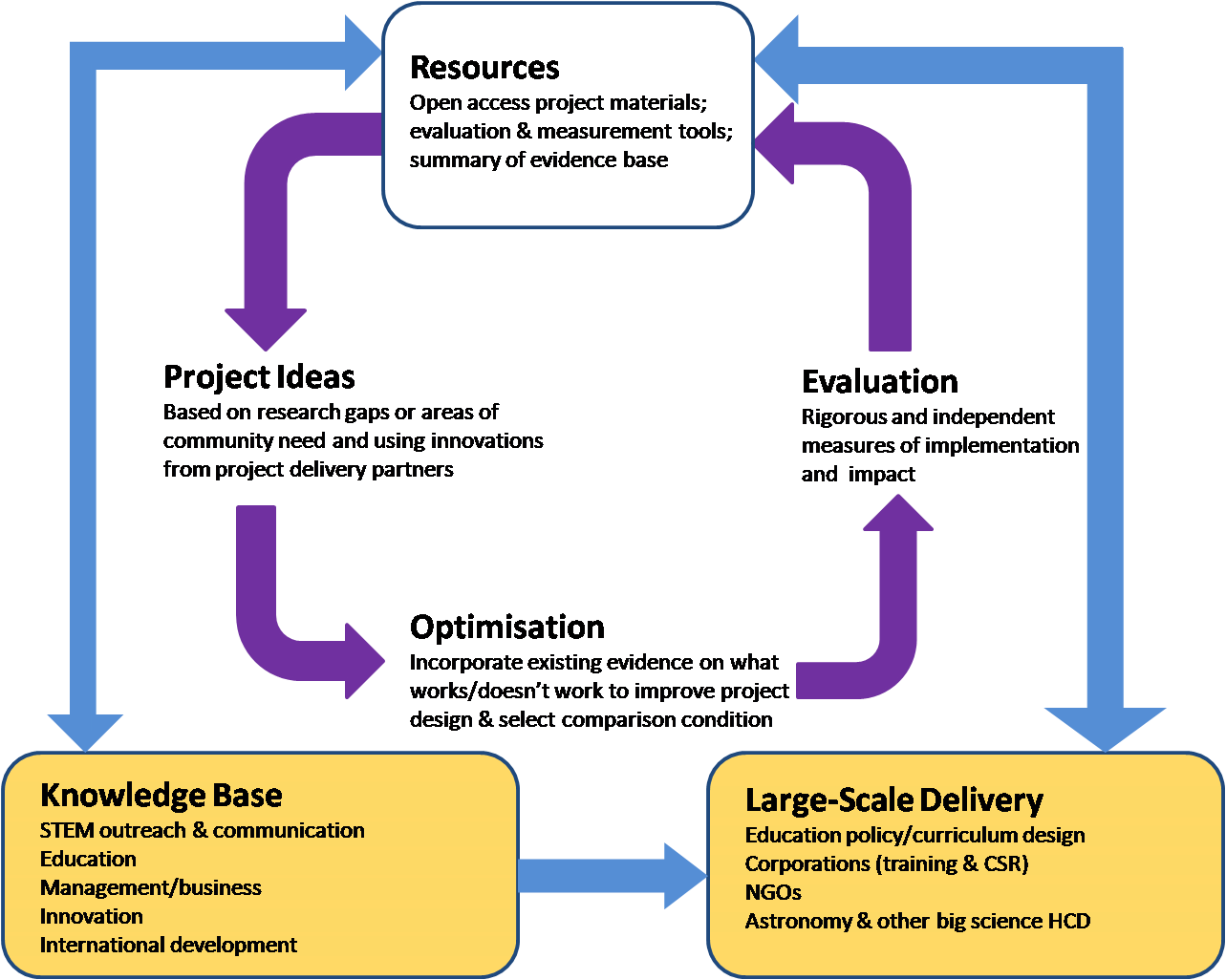}
    \caption{OAD Impact Cycle}
    \label{impactcycle}
\end{figure}

The philosophy of the OAD is to build developmental interventions on what is known to work, improve those projects over time, and to avoid interventions that have unintended negative consequences. A simple illustration of how this works is shown in Fig. \ref{impactcycle}, this figure summarises the OAD impact cycle. The cycle is a "map" through which future projects can:
\begin{itemize}
    \item Access OAD resources and available scientific evidence on effective science communication, education and international development strategies.
    \item Use these resources to develop scientifically-informed project designs
ensure they are not simply replicating past projects.
    \item Draw from past projects’ lessons when it comes to planning and designing their projects, thus avoiding the repetition of mistakes.
    \item Monitor their project’s implementation, providing a foundation to attract funding, increase the project’s scale and enable other projects to replicate their success.
    \item Evaluate their project’s impact, thus contributing to a growing scientific evidence base on what works and providing evidence of how astronomy can most effectively contribute to development.
\end{itemize}

\section{The Impact Cycle}
The impact cycle begins with a project idea (Stage 1), which is then turned into a project design through optimisation (Stage 2). During and after the project execution, evaluation is implemented to help understand the impact of the project and adds to the evidence base for future projects. 

\subsection{Resources}\label{resources}
Providing accessible resources plays a crucial role in supporting the design and implementation of effective astronomy-for-development projects. A number of resources are available to assist those interested in this work, including evidence, interventions, a project toolbox, data, and partners. Each of these key points is discussed in this section.
\begin{itemize}
    \item \textbf{Evidence}\linebreak
    The resources include guidance on finding and interpreting relevant evidence, such as understanding research design and validity. It features a collection of systematic reviews and brief evidence appraisals generated by the community, partners, the OAD, and external contributors (where open access). Additionally, it provides short best-practice info sheets or an intervention catalogue, particularly focusing on mechanisms like self-fulfilling prophecies, bias, and incentives. The resources also include links to relevant websites, review collections, and research articles from platforms such as the Campbell Collaboration, 3ie, Better Evaluation, and the OECD, along with curated lists of books and articles on education, outreach, and development. Finally, it references policy reports and journal content, including research articles, commentaries, and editorials on evidence-based outreach, education, and development.

    \item \textbf{Interventions}\linebreak
    The resources feature a comprehensive catalogue of all OAD projects, detailing target populations, intended outcomes, identified problems, and corresponding evidence appraisals. It also includes a repository of project manuals with sufficient detail to support replicability. Each manual provides a review of the underlying theory and evidence base, a description of key components (including delivery agents, locations, duration, intensity, and activity schedules), as well as supporting materials such as worksheets, exercises, slides, and assessments. Criteria for participant and staff selection or training is also outlined, alongside links to relevant data. Additionally, the resource includes links to replicable materials from other sources, such as NASA, each connected to corresponding evidence appraisals.

     \item \textbf{Toolbox and Training}\\
     The resources include comprehensive guidance for each stage of project implementation, from design to dissemination. Under \textbf{Project Design}, tools are available such as the PICO framework for evidence-based planning, guidance on developing a Theory of Change, methods for estimating cost-to-impact ratios, and resources on project management, budgeting, and equity mainstreaming. The \textbf{Assessments} section features a collection of validated tools for measuring common learning outcomes, along with guidance on developing new assessments and collecting data on broader behavioural outcomes such as school completion. The \textbf{Evaluation} component, linked to the evaluation stage of the project cycle, offers detailed frameworks, design templates, and practical advice, including consent and information form templates, ethical approval procedures, data collection and monitoring tools, and software for random assignment. Finally, the \textbf{Scale-Up, Publication, and Dissemination} section, connected to the \textit{large-scale delivery phase}, will provide resources on manualisation, publication, fundraising, and publicity to support the broader impact and sustainability of projects.
     
     \item \textbf{Data}\linebreak
    The resources include a comprehensive collection of data to support project planning, evaluation, and learning. Under \textbf{Needs Analyses} (linked from section \ref{project ideas}, "Project Ideas"), it features survey results, such as assessments of astronomy capacity in regions like West Africa, and on-the-ground reports collected through Regional Offices of Astronomy for Development (ROADs) and other local networks. The \textbf{Evaluation Data} section (linked from section \ref{evaluation}, "Evaluation") contains primary project impact evaluation reports, individual-level data from these evaluations, and follow-up contact information for continued engagement with project participants. Finally, the \textbf{Implementation Data} section providing detailed information on project delivery, including the number of participants recruited or reached (both intended and actual), cost per person, qualitative and process data, evaluations of specific component mechanisms, and documentation of delivery challenges.

     \item \textbf{Partners}\linebreak
     The resources include a curated database of potential partners and collaborative opportunities to strengthen project implementation and impact. Under \textbf{Development Partners}, it features a list of organisations and collaborators from the development sector who can provide expertise or support in project execution. The \textbf{Development Projects} section showcases initiatives that users can engage with or help disseminate, for example, integrating tested positive parenting activities into astronomy outreach events. Lastly, the \textbf{Astronomy Partners} section highlights astronomy-related programmes and networks, such as Universe Awareness (UNAWE) and Galileo Teacher Training Program (GTTP), that can serve as valuable collaborators for joint activities and knowledge exchange.
    \end{itemize}

\subsection{Project Ideas}\label{project ideas}
The second step of the OAD Impact Cycle is Project Ideas. As mentioned, the OAD Annual Call for Proposals follows a two-stage selection process. Stage 1 of this process involves the review of project ideas. This step comprises two key components: the call structure and the selection of projects. This section provides more details.

\begin{itemize}
    \item \textbf{Call Structure}\\
    The goal is to expand the dissemination beyond astronomy. To achieve this, targeted information on areas requiring intervention is provided. This includes identifying gaps in evidence, such as tested interventions needing replication in new contexts, or promising and widely used mechanisms that require further evaluation. The areas of need are highlighted, including priority development issues like socio-economic or gender inclusion, literacy, or effective interventions that require additional support, with links to relevant resources. The application process encourages greater specificity and careful consideration of the problem, intervention features, comparison conditions, and expected outcomes. A two-stage process is implemented, where Stage 1 focuses on generating ideas and Stage 2 emphasizes optimisation of the proposed interventions.

    \item \textbf{Selections}\linebreak
    The selection process involves anonymous submissions and is guided by explicit criteria to ensure quality and relevance. Proposals are assessed based on whether they clearly specify the PICO framework (as outlined in section \ref{optimisation}, “Optimisation”), align with existing evidence, pose low risk of harm, and are feasible to implement. Additional considerations include positive cost-to-impact estimates, scalability, evaluability (or justification if evaluation is not required), and the extent to which the project addresses priority outcomes.
\end{itemize}

\subsection{Optimisation}\label{optimisation}
The third step of the OAD Impact Cycle is Optimisation. Projects selected during Stage 1 undergo the optimisation process. During this stage, project ideas are refined and developed into a full project design. This section provides more information on what the optimisation stage involves.

\begin{itemize}
    \item \textbf{PICO}\linebreak
    The OAD works with projects to explicitly specify and refine the \textbf{P}roblem(s), \textbf{I}ntervention, \textbf{C}omparison, and \textbf{O}utcome(s) using the PICO tools from the Resources Toolbox - discussed in section \ref{resources}. Projects addressing the same problem and aiming to achieve similar outcomes are connected, and, where possible, prior OAD interventions or “usual conditions” are used as comparisons, with links to past project consolidation provided in the Resources section.
    
    \item \textbf{Theory and Outcomes}\\
    The OAD and Task Forces actively support projects in developing their theory and outcomes. Guidance is provided on using the Theory of Change from the Resources Toolbox, help projects build evidence-informed theories, identify which aspects of the theory need evaluation, and specify measurable and theoretically relevant outcomes. In addition, projects incorporate equity mainstreaming and budgeting by considering gender, race, socio-economic status, and (dis)ability perspectives, using these considerations to optimise both project design and budgeting.

    \item \textbf{Monitoring and Evaluation}\\
    Monitoring and evaluation are integrated into the project proposal, design, and implementation processes. This includes using online project registration forms to collect data on participants such as teachers, students, and schools, standardised online financial and implementation reporting, and an automated system for flagging planned milestones from proposal data. Evaluation plans and timelines are built into the project design by using the Evaluation Framework and Resources Toolbox to design and plan evaluations at the project design stage, and by incorporating evaluation, data collection, and reporting into both budget and time planning \cite{monitoring and evaluation}.
\end{itemize}

\subsection{Evaluation}\label{evaluation}
Evaluation is a critical component of the OAD Impact Cycle, drawing on resources, project ideas, and the optimisation process to assess and enhance project effectiveness. This section highlights the Evaluation Framework, outlines different types of evaluation, and provides illustrative examples to guide practice. It also details the support available to projects, helping teams design robust evaluation plans, collect and analyse data, and use findings to inform decision-making, improve outcomes, and share lessons learned \cite{monitoring and evaluation}.

\begin{itemize}
    \item \textbf{Framework}\linebreak
    The evaluation framework is designed to be theory-informed, feasible, and rigorous. It uses theory to address unanswered questions about project mechanisms and impacts, guiding the selection of appropriate outcome or impact measures. Evaluations are designed to be feasible by leveraging existing data where possible, using small sample sizes if necessary, enabling follow-up while planning for short-term assessment, and minimising the burden on project leaders through tools such as automated registration and random assignment. At the same time, evaluations maintain rigor by selecting the design best suited to answer the questions of interest, such as using random assignment to measure causal impacts, conducting pre-trial registration, adhering to CONSORT guidelines in post-trial reporting, and seeking partners with relevant experience, with additional guidance available in the Resources section.

    \item \textbf{Types of Evaluation}\\
    Evaluation can take several forms, each serving a distinct purpose. \textbf{Needs analysis} involves using random sampling or respondent-driven survey research, for example, to measure a university’s gender disparities, or analysing open-access data, with guidance available in the Resources section. \textbf{Feasibility evaluation} tests the practicality and cost of delivering an innovative intervention, such as remote telescopes, or a planned data collection strategy, like response rates for email follow-ups. \textbf{Process evaluation} explores causal mechanisms through experimental approaches and captures participant and project delivery experiences using qualitative methods. \textbf{Impact evaluation} assesses whole-project outcomes using experimental designs, applying a hierarchy of evidence and random assignment at the appropriate level of analysis, such as school, classroom, or individual, while recognising that long-term follow-up may require additional funding.

    \item \textbf{Examples of Evaluation Types}\\
    \begin{enumerate}
        \item \textbf{Needs Analysis}
        Needs analysis evaluation addresses questions about the necessity and context of astronomy education, such as whether astronomy should be included in the curriculum or if it would replace existing content, what students have already learned, and whether outcomes differ by gender, race, socio-economic status, or (dis)ability. It also considers which skills are most in demand in local labour markets and identifies barriers that may cause students, or specific subsets of students, to underperform or drop out of science degrees or careers. Evaluation designs for needs analysis typically combine methods such as analysis of public data (e.g., TIMSS, PERL, World Bank indicators, local education data), key informant interviews with students, teachers, and administrators, documentary review of curriculum materials and assessment instruments, and surveys using random sampling or respondent-driven sampling strategies to obtain representative estimates.
        \item \textbf{Feasibility Evaluation}
        Feasibility evaluation helps determine whether a project can be practically and effectively implemented in a given context. Example questions include: Are internet connections stable and fast enough for teachers in Francophone West Africa to participate in an online teacher training course? Will students in Central America be willing to travel 15 hours by bus to attend a two-week astronomy workshop? Can an Ultrascope be built feasibly and remain within budget in South Africa? The evaluation design involves collecting monitoring data on project design, such as how projects were advertised, where, and for how long; monitoring data on project uptake, including the number of student applications and offers of volunteer support; budget data, such as the cost of components and time required for construction; and implementation data, capturing how the project was carried out in practice, unplanned adaptations that occurred on the ground, and components that could not be delivered.
        \item \textbf{Process Evaluation}
        Process evaluation focuses on understanding how a project is implemented and how participants experience it. Example questions include: Did students’ science test results improve because they were more interested in the course? Were there any unanticipated positive or negative consequences for participants? How did participants feel about the project? How could the project be improved? Did the gender balance of role models in a career brochure affect learning outcomes? Evaluation designs for process evaluation vary depending on the question. If a causal mechanism is specified, experimental approaches, such as random assignment to materials with different proportions of male or female role models, can measure the short-term effects of specific project components. For other questions, qualitative methods like open interviews, focus groups, and post-project surveys are used to capture participants’ perspectives and insights.
        \item \textbf{Impact Evaluation}
        Impact evaluation assesses whether a project “works,” that is, whether it causes changes in the target outcome that would not have occurred in its absence. Evaluation designs follow an evidence hierarchy and measure causal impact relative to a comparison condition (A vs. B). Exposure to the project must be statistically independent of the outcome, typically achieved through random assignment, though evaluating overall impact may not always be possible. For example, in an over-subscribed workshop, all eligible applicants (the population of interest) can be randomly assigned either to attend the workshop (A) or to access an online course (B). Pre- and post-workshop knowledge is then measured, and a statistical comparison of A versus B determines the size of the workshop’s impact relative to the alternative, with cost-effectiveness also calculated.
    \end{enumerate}
    \item \textbf{Support}
    Evaluation support from the OAD plays a critical role in helping project leaders design and implement effective assessments. The OAD provides guidance and resources to assist teams in planning appropriate evaluation approaches, drawing on internal expertise as well as relevant external partners with specialised experience. This support includes advice on selecting suitable evaluation designs, identifying measurable outcomes, developing data collection and analysis plans, and ensuring adherence to ethical and methodological standards. Project leaders are also guided in using available tools and templates, accessing evidence-informed practices, and connecting with collaborators who can provide technical or sector-specific input. Further resources and detailed guidance are accessible through the OAD “Resources” section.
\end{itemize}

\subsection{Knowledge Base}
The knowledge base is a central repository of evidence, insights, and practical guidance that supports effective astronomy-for-development projects. It brings together resources, project ideas, optimisation processes, and evaluation data to create a comprehensive understanding of what works, for whom, and under what conditions. Resources provide foundational evidence and tools, while project ideas contribute emerging concepts and innovative approaches. The optimisation stage refines these ideas into well-designed, evidence-informed projects, and evaluation generates data on their outcomes and impact. Together, these elements continuously feed into the knowledge base, ensuring it remains up-to-date, practical, and relevant for future projects and stakeholders.
\begin{itemize}
    \item \textbf{Target Communities}\\
    The knowledge base serves as a resource for a wide range of stakeholders across multiple sectors. In \textbf{STEM outreach and communication}, academic research, practitioners, and supporting institutions can access insights to inform their work. In \textbf{education}, academic researchers, teachers at all levels, teacher training institutions, government departments, and non-governmental organisations (NGOs) can benefit from findings on effective educational interventions - for example, investigating whether incorporating astronomy into mathematics lessons improves engagement or learning outcomes. In \textbf{management and business}, academic research, human resources in industry and academia, and corporate social responsibility initiatives can draw on evidence to guide practices. For \textbf{monitoring and evaluation}, professional and academic evaluation specialists, government departments, and multilateral organisations can utilise the knowledge base, while in \textbf{development economics}, academic researchers, development banks, and NGOs can access relevant research to support evidence-informed decision-making.

    \item \textbf{Publications}\linebreak
    The knowledge base includes a wide range of publications, both open-access through the OAD website and in peer-reviewed journals. These publications enable OAD-generated evidence to be incorporated into systematic reviews and policy analyses, contributing to the global evidence base on science for development. Publications may be prepared by project leaders, covering areas such as needs analyses, process evaluations, theory, and project manuals; by the OAD, including process evaluations, impact evaluations, systematic reviews, evidence appraisals, and meta-analyses that combine impact evaluations; or by external researchers and partners, who analyse primary and evaluation data. The OAD and its partners support project leaders in publishing their work, while impact evaluations and some process evaluations are published independently by the OAD or other external evaluators to ensure objectivity. Some publications from the OAD include: \cite{comment1}-\cite{narratives} and from project leaders and partners \cite{resource4}-\cite{comment4}.

    \item \textbf{Policy Briefs}\\
    The knowledge base supports the development of policy briefs by drawing on accumulated insights from needs analyses and evaluations, providing a platform for informing science and human capital development policymakers. Policy briefs serve as tools for translating “lessons learned” into practice, offering concise summaries of evidence related to alternative policies. Resources provided by the OAD through the Impact Cycle, particularly project materials, data, and publications, are made available for use in policy briefs prepared by external stakeholders, including think tanks, governments, science bodies, lobbyists, unions, and NGOs.
\end{itemize}

\subsection{Large-Scale Delivery}
Large-Scale Delivery represents the stage where astronomy-for-development projects move beyond pilot or small-scale implementation to achieve broader, sustainable impact. It builds on the foundations established in earlier stages of the OAD Impact Cycle, drawing on the \textbf{Resources} that support evidence-based design, the \textbf{Project Ideas} generated through community input, the \textbf{Optimisation} processes that refine interventions, and the \textbf{Evaluation} findings that provide feedback and proof of effectiveness. Insights from the \textbf{Knowledge Base} further strengthen this stage by ensuring that scaling efforts are grounded in research, experience, and best practices. The target audience for large-scale delivery includes policymakers, industry partners, NGOs, and volunteers, who play a vital role in supporting, adopting, and expanding successful astronomy-for-development initiatives to reach wider communities and create lasting change.

\begin{itemize}
    \item \textbf{Policymakers}\linebreak
    In large-scale delivery, policymakers play a crucial role in driving systemic change by using evidence to inform decisions on national curricula, science policies, and educational funding streams. Through the OAD Impact Cycle and its supporting Resources (discussed in section \ref{resources}), relevant evidence is generated and made accessible to guide choices between alternative priorities and programmes. This includes evidence of impact and cost-effectiveness demonstrated through project evaluations, as well as targeted dissemination of findings via policy briefs and evidence reviews, ensuring that decision-making is grounded in robust, data-driven insights.
    
    \item \textbf{Industry}\linebreak
    Collaboration with industry plays a key role in expanding the impact of astronomy-for-development initiatives. Evidence gathered through the OAD’s work demonstrates that astronomy can be an effective tool for addressing labour-market skills shortages in fields such as data science, programming, engineering, and software development. Industry engagement, through investment in skills training workshops, contributions to higher education curriculum development, and the creation of scholarship programmes, can enable large-scale and sustainable expansion of related projects, ensuring that astronomy continues to contribute meaningfully to workforce development and innovation.

    \item \textbf{NGOs}\linebreak
    Large-scale delivery through NGOs involves collaboration with major international non-governmental organisations such as Save the Children and Oxfam, as well as governmental departments like Department for International Development (DFID) and United States Agency for International Development (USAID), and multilateral organisations including UNICEF, UNDP, and the World Bank. These institutions support educational and human capital development programmes on a much larger scale than the OAD can achieve alone. While these organisations typically design and implement their own programmes, they are often willing to adopt approaches that have been demonstrated to be effective and cost-efficient. The outputs of the OAD Impact Cycle, such as evidence-based findings and replicable project materials, provide these agencies with the necessary information and resources both to be convinced of the value of OAD project approaches and to scale them up for broader impact.

    \item \textbf{Volunteers}\linebreak
    Many science outreach and development projects are designed and delivered by volunteers, who play a crucial role in extending the reach and impact of astronomy-for-development initiatives. Although numerous astronomy outreach and educational resources are available online, there is currently little strategic coordination to help volunteers (and teachers) select those most suitable for their needs and most likely to be effective. Through the OAD Impact Cycle, volunteers can easily identify interventions that have been shown to be effective and replicate these proven approaches, enabling more consistent, evidence-based, and impactful delivery of astronomy-related activities across different contexts.
\end{itemize}

\section{Conclusion}
The OAD Impact Cycle provides a structured approach to designing, implementing, and evaluating astronomy-for-development projects. By integrating resources, project ideas, and optimisation, the cycle ensures that projects are evidence-informed, feasible, and responsive to community needs. Curated resources, including evidence reviews, project manuals, intervention tools, data, and partner networks, support all stages of project planning and delivery. The evaluation framework promotes theory-informed, rigorous, and practical assessment, encompassing needs analysis, feasibility, process, and impact evaluation. Through guidance, templates, and partnerships, the OAD helps project leaders define measurable outcomes, incorporate equity considerations, and assess causal impacts, ensuring that interventions are scalable, cost-effective, and impactful. Overall, this cycle fosters the systematic development of astronomy-for-development initiatives that generate measurable social and educational benefits while supporting evidence-based decision-making.

\end{document}